\documentclass[aps,prc,twocolumn,nofootinbib]{revtex4}

\usepackage{graphicx}

\newcommand{\be}{\begin{equation}}
\newcommand{\ee}{\end{equation}}
\newcommand{\bea}{\begin{eqnarray}}
\newcommand{\eea}{\end{eqnarray}}
\newcommand{\mybibitem}{\bibitem}

\newcommand{\gton}{\mathrel{\lower.9ex \hbox{$\stackrel{\displaystyle 
>}{\sim}$}}} 
\newcommand{\lton}{\mathrel{\lower.9ex \hbox{$\stackrel{\displaystyle 
<}{\sim}$}}}

\newcommand{\vb}{{\bf b}}

\newcommand{\vx}{{\bf x}}

\newcommand{\vp}{{\bf p}}

\begin{document}

\title{How AMPT generates large elliptic flow with small cross sections}

 \author{Denes Molnar}
 \affiliation{Department of Physics and Astronomy,
              Purdue University, West Lafayette, IN 47907}

\date{\today}

\begin{abstract}
We resolve the long-standing open question of how the transport model AMPT\cite{Xu:2011fe} 
manages to generate sufficiently high
elliptic flow ($v_2$) in A+A reactions with only few-millibarn $2\to 2$ partonic cross sections - in apparent
contradiction with an early study by Molnar and Gyulassy\cite{v2}. Through detailed comparisons
with the covariant 
Molnar's Parton Cascade (MPC), we pinpoint which features of initial conditions,
interactions, and dynamics encoded in the partonic stage of AMPT allow it to circumvent the ``opacity puzzle'' at RHIC.
\end{abstract}

\maketitle

\section{Introduction}

AMPT\cite{AMPT}, which stands for ``A Multi-Phase Transport'', is a 
highly successful model for high-energy nuclear reactions that describes
reasonably well a wide range of observables at RHIC and LHC energies.
Given the complexity of the model, it is natural to yearn for some simple
understanding of what makes it work so well. 
AMPT is particularly interesting because it is rather different from relativistic hydrodynamics.
For example, elliptic flow from AMPT is largely governed by the anisotropic probability
for particles to escape the collision zone\cite{aniescape}.
Also, with the $2\to 2$ scattering
and moderate opacities $N_{coll}\sim {\cal O}(5)$ involved, the transport evolution in AMPT is far
from the hydrodynamic limit\cite{hytrv2}.

Nevertheless, AMPT can generate sizeable
collective effects in A+A collisions at RHIC, such as elliptic flow\cite{Xu:2011fe,Lin:2014tya}.
In fact, it does that with only few-millibarn elastic parton-parton 
scatterings.
This has been an open question for a long time because it 
is at loggerheads with an earlier study by Molnar and Gyulassy\cite{v2}, which
found that elastic parton cross sections $\sigma \sim 40-50$~mb are required to reach the charged hadron $v_2$
data.
Radiative $2\leftrightarrow 3$ scatterings would help provide a way
out\cite{BAMPSv2} but AMPT manages with only $2\to 2$ scatterings.

Here we provide a resolution to the puzzle.
First we highlight the main differences between AMPT and the study in Ref.~\cite{v2} (Sec.~\ref{Sc:comp}).
Then, through detailed comparisons with covariant Boltzmann transport solutions,
we identify key features that allow AMPT to generate high elliptic flow with
small cross sections (Secs.~\ref{Sc:prof} and \ref{Sc:mom}).
The issue of noncovariant dynamics due to the naive cascade algorithm in AMPT is also addressed (Sec.~\ref{Sc:repro}).

\section{Covariant transport theory}
\label{Sc:transport}

We utilize on-shell covariant transport theory
to perform detailed comparisons against AMPT.
For a system with elastic
$2\to 2$ interactions (see, e.g., Refs. \cite{Molnar:2000jh,MolnarWolff}), 
the evolution of
the phase space density $f$ of species $i$ is given by the nonlinear Boltzmann transport
equation (BTE)
\be
p^\mu \partial_\mu f_i(x,\vp) = S_i(x,\vp) +
\sum_j C_{ij}[f_i,f_j](x, \vp) \ .
\label{BTE}
\ee
Here the source term $S_i$ encodes the initial conditions for species $i$,
while the two-body collision term is
$$
C_{ij}(x,\vp_1) 
\equiv \int\limits_2 \!\!\!\!\int\limits_3 \!\!\!\!\int\limits_4
\left(f_{i3} f_{j4} - f_{i1} f_{j2}\right)
\, \bar W_{12\to 34}^{ij\to ij}  \, \delta^4(12 - 34)
$$
with shorthands
$\int\limits_a \equiv \int d^3p_a / (2 E_a)$, 
$f_{ia} \equiv f_i(x,\vp_a)$, and
$\delta^4(ab - cd) \equiv \delta^4(p_a + p_b - p_c - p_d)$.
The transition probability $\bar W_{12\to 34}^{ij\to ij}$ for two-body $ij\to ij$ scattering
with momenta 
$p_1 + p_2 \to p_3 + p_4$ is given by the differential cross section as
$\bar W^{ij\to ij} = 4 s d\sigma^{ij\to ij} / d\Omega_{cm}$, where $s \equiv (p_1 + p_2)^2$ 
is the usual Mandelstam variable, and the
solid angle is taken in the c.m. frame of the microscopic two-body collision.

To compute covariant transport solutions numerically, 
we utilize the MPC/Cascade algorithm from Molnar's Program Collection (MPC) \cite{Molnar:2000jh,MPCcode}. 
The algorithm implements the BTE collision term through particle scattering at closest approach 
(in the two-body c.o.m. frame).
Covariance is ensured\cite{Zhang:1998tj,Molnar:2000jh} via particle
subdivision 
$f_i \to \ell f_i$, $\sigma_{ij} \to \sigma_{ij}/\ell$,
which is an exact scaling of the BTE (\ref{BTE}).
For large $\ell$, the interaction range $d = \sqrt{\sigma/\pi}$ shrinks as $\ell^{-1/2}$, 
restoring locality and covariance. 
While different transport codes in general give different, noncovariant results
when no subdivision is used, the results all converge to the same solution at high enough subdivision\cite{Cheng:2001dz}.

Similarly to MPC, Zhang's Parton Cascade (ZPC)\cite{ZPC} embedded in AMPT can also employ particle subdivision, in principle. However,
in practice, AMPT runs ZPC without subdivision (i.e., with $\ell = 1$). 
Noncovariant artifacts in AMPT will be investigated in Sec.~\ref{Sc:repro}.

\section{Step-by-step AMPT vs MPC comparisons}
\label{Sc:comp}

First we discuss the main differences between the partonic stage of AMPT and the MPC study in Ref.~\cite{v2}.
Then, a series of MPC simulations are presented that are gradually more and more ``AMPT-like'',
until we pinpoint eventually how AMPT generates large elliptic flow with small cross sections.

\subsection{AMPT vs old MPC study}
\label{Sc:diff}

The opacity vs elliptic flow study by Gyulassy and Molnar~\cite{v2} was for massless gluons,
with Debye-screened perturbative QCD differential elastic cross section 
\be
\frac{d\sigma_{gg}}{dt} = \sigma_{gg} \left(1+\frac{\mu^2}{s}\right) \frac{1}{(t - \mu^2)^2} \ ,
\quad \sigma_{gg} = \frac{9\pi\alpha_s^2}{2\mu^2} \ .
\label{dsigmadt}
\ee  
Here, $s$ and $t$ are the usual Mandelstam variables, $\alpha_s$
is the strong coupling, $\mu$ is the Debye screening mass, and $\sigma_{gg}$ is the total cross section.
The initial conditions for Au+Au at RHIC at $b=8$~fm were longitudinally boost-invariant,
with locally thermal momenta at temperature 
$T_0 = 0.7$~GeV at fixed Bjorken proper time $\tau_0 = 0.1$~fm,
a flat coordinate rapidity profile $dN_g/d\eta(b=8~{\rm fm}) \equiv n_\eta \approx 240$ in a wide
$|\eta| < \eta_0$ window with $\eta_0 = 5$ (``Bjorken sausage''), 
and a binary collision transverse profile for two Woods-Saxon nuclei. This corresponds to
the joint momentum rapidity ($y$), transverse momentum ($\vp_T$), coordinate rapidity ($\eta$), and 
transverse position $\vx_T$ distribution
\bea
\frac{dN_g(\tau = \tau_0)}{dy d^2 p_T d\eta d^2 x_T}
&=& \frac{n_\eta p_T \, \cosh \xi}{8\pi T_0^3}\, \Theta(|\eta| < \eta_0) 
\nonumber \\
&& \qquad \times f(\vx_T)\, e^{-\frac{p_T \cosh \xi}{T_0}}
\label{thermal_boostinv}
\eea
with $\xi \equiv \eta - y$,
\be
f(\vx_T) = \frac{T_A(\vx_T - \vb/2) T_A(\vx_T + \vb/2)}{\int d^2 x_T T_A(\vx_T - \vb/2) T_A(\vx_T + \vb/2)}
\ee
where $\vb$ is the impact parameter vector, and 
\be
T_A(\vx_T) \equiv \int dz \rho_A(\sqrt{z^2 + \vx_T^2}) 
\ee
is the thickness function for nucleus $A$.
The Debye mass was set to the initial temperature $\mu = T_0$, and the elastic
cross section was then varied by dialing $\alpha_s$. Large $\sigma_{gg} \sim 40-50$~mb
were found necessary to reproduce pion $v_2(p_T)$ in Au+Au at RHIC\cite{v2}.

In contrast, in its partonic stage AMPT simulates a system made of quarks 
of nonzero mass $m_u = 9.9$~MeV, $m_d = 5.4$~MeV, $m_s=199$~MeV.
The parton transport evolution is computed using Zhang's ZPC\cite{ZPC},
with elastic scattering cross sections of the same form (\ref{dsigmadt}) as in Ref.~\cite{v2} 
but applied flavor-independently to quarks 
(i.e., AMPT ignores
the drop in the Casimir for quark-quark scattering relative to gluon-gluon scattering). 
Typically, $\alpha_s = 0.33$ is set,
and the elastic cross section is then controlled by adjusting the Debye mass - the default value 
$\mu \approx 0.45$~GeV gives $\sigma_{qq} = 3$~mb. Initial conditions for the partons
 are generated via the ``string melting''
mechanism\cite{AMPTsm} that converts 
strings formed via HIJING\cite{HIJING} back into quarks (the strings are first decayed
to hadrons that are then converted to their quark constituents).
This results in nonuniform initial rapidity distributions, nonthermal momenta, a broad distribution
of quark formation times,
and a variety of specific correlations such as
narrower than thermal $\eta-y$ correlations, and anticorrelation between formation time and $p_T$ (reflecting
$\tau \sim 1/p_T$ formation physics). AMPT also incorporates event-by-event fluctuations in nucleon positions,
which affect string formation, and thus lead to event-by-event fluctuations in the initial quark distributions, 
such as the transverse position and total number of quarks in the event.

Clearly, there are many differences between AMPT and the old MPC study in Ref.~\cite{v2}. 
In what follows, we pursue MPC simulations that are systematically more and more AMPT-like. 
For simplicity, we limit the investigation here to Au+Au collisions at top RHIC energy 
$\sqrt{s_{NN}} = 200$~GeV, 
at fixed impact parameter $b = 8$~fm.
For AMPT, we analyze 90000 events obtained using AMPT version 2.26t5 \cite{AMPTcode};
while in each iteration with MPC, we generate statistics equivalent to 20000 events (i.e., $N_{events} \ell = 20000$) 
with MPC version 1.9 \cite{MPCcode}.
 
{\em First of all, in the elastic cross section $d\sigma/dt$
we set $\alpha_s = 0.33$ and $\mu = 0.447$~GeV, which match the default AMPT parameters for $\sigma_{qq} = 3$~mb.}
This makes the microscopic parton dynamics in MPC and AMPT identical, provided AMPT solves the covariant
BTE.

\subsection{Transverse profile and \boldmath{$dN/dy$}}
\label{Sc:prof}

\begin{figure}[h]
\leavevmode
\hspace*{-1mm}\includegraphics[height=50mm]{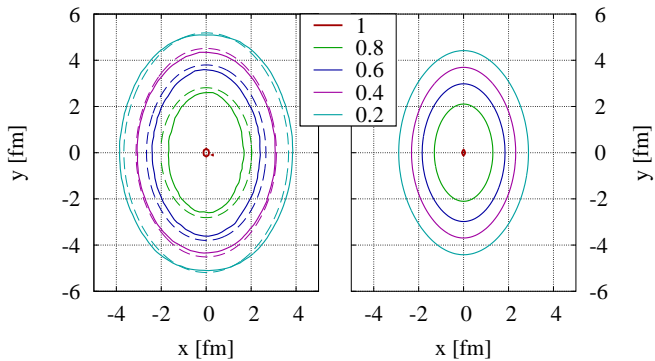}
\caption{Initial quark density profile in the transverse plane
at midrapidity ($|y| < 1$) for Au+Au at RHIC at $b=8$~fm from AMPT (left panel), compared
to the binary collisions profile of initial minijet gluons used by Ref.~\cite{v2} (right panel).
For both distributions, normalized contour lines are shown (solid lines) 
at 1, 0.8, 0.6, 0.4, and 0.2 times the maximum of the 
respective distribution. The $x$ direction is the impact parameter direction.
Dashed lines in the left panel indicate similarly normalized contours for the
wounded nucleon profile.
}
\label{Fig:comp_dNdxdy}
\end{figure}

Next we turn to the initial transverse profile and rapidity density.
Figure~\ref{Fig:comp_dNdxdy} compares the initial (event-averaged) quark transverse 
density profile at midrapidity $|y| <1$ from AMPT (left panel, solid curves), 
to that for the study by Ref.~\cite{v2} (right panel).
Clearly, the binary collision profile for minijets is significantly more compact than the distribution from AMPT.
In fact, the AMPT profile is rather close to the wounded nucleon profile for Au+Au (left panel, dashed): 
\be
\frac{dN}{d^2 x_T} \propto T_A(\vx_T - \vb/2) (1- e^{-\sigma_{NN} T_A(\vx_T + \vb/2)}) + (\vb \to -\vb) \ ,
\ee
where $\sigma_{NN}$ is the inelastic nucleon-nucleon cross section ($\approx 42$~mb at top RHIC energy).
{\em Therefore, our second adjustment 
from here on is to always initialize MPC with the wounded nucleon profile for Au+Au at RHIC with $b=8$~fm.}
Compared to the binary collision shape,
the initial spatial eccentricity is about $15$\% smaller ($\varepsilon \approx 0.35$ versus 0.30 only), 
which also reduces the final elliptic flow by a similar
portion compared to the original calculations in Ref.~\cite{v2}.
%

\begin{figure}[h]
\leavevmode
\includegraphics[height=65mm]{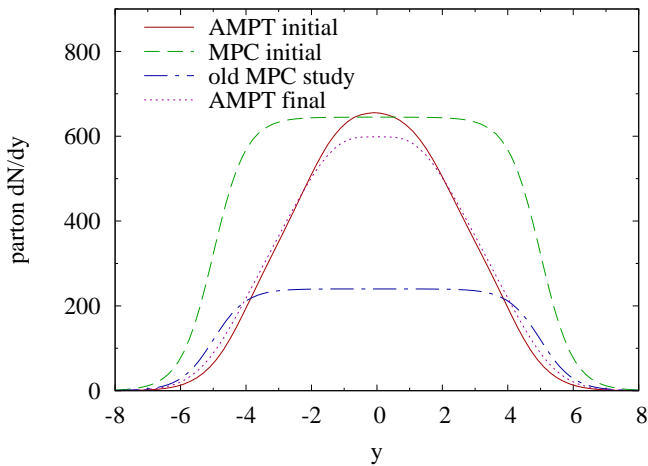}
\caption{Initial quark rapidity distribution $dN/dy$ in Au+Au at RHIC with $b=8$~fm
from AMPT (solid line), compared to that for the covariant MPC simulations in this work (dashed) and 
to the initial gluon $dN_g/dy$ in the study by Ref.~\cite{v2} (dashed-dotted). 
The quark rapidity distribution at the end of the AMPT partonic stage is also shown (dotted).
}
\label{Fig:comp_dNdy}
\end{figure}

Figure~\ref{Fig:comp_dNdy} contrasts the initial rapidity distributions. The quark distribution in AMPT (solid curve) 
has a roughly triangular shape that reaches $dN/dy \approx 645$ at midrapidity, whereas the initial
conditions in Ref.~\cite{v2} correspond to an about $2.5\times$ lower and roughly flat $dN/dy \approx 240$ in
a window of nearly $\Delta y \pm 4$ near midrapidity%
\footnote{In the longitudinally boost-invariant scenario, $dN/dy$ closely tracks
the box-shaped coordinate rapidity $dN/d\eta$ distribution, except for a smearing by about $\pm 1$ units
near the $y \approx \pm \eta_0$ edges
due to the thermal $\eta-y$ correlation encoded in (\ref{thermal_boostinv}).
}.
{\em Our third adjustment for MPC simulations, therefore, is to increase the density to $dN/dy = 645$},
via setting $n_\eta = 645$ (dashed line).
One could also include the AMPT rapidity shape in MPC but we prefer to keep things as simple as possible. 
The final quark rapidity density from AMPT (dotted) is almost the same as the initial $dN/dy$ (solid),
indicating that the rapidity distribution changes very little during the evolution.
%

\begin{figure}[h]
\leavevmode
\hspace*{-1mm} \includegraphics[height=63mm]{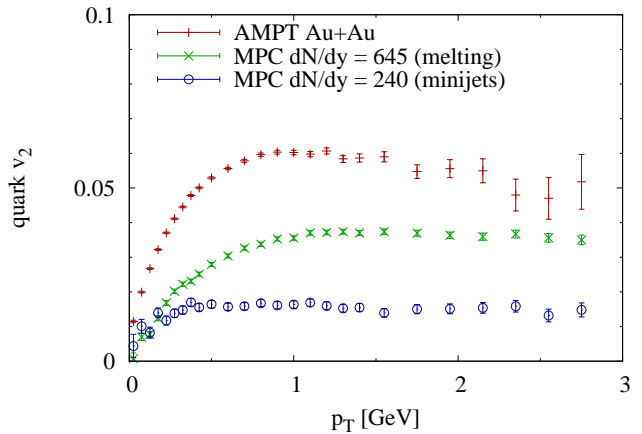}
\caption{Differential quark elliptic flow $v_2$ vs $p_T$ at midrapidity ($|y| < 1$) in Au+Au at RHIC with $b=8$~fm
from AMPT (pluses), compared to covariant results from MPC using the same partonic $d\sigma/dt$ as in AMPT with $\sigma_{qq} = 3$~mb. MPC calculations from longitudinally boost-invariant initial
conditions are shown for $dN/dy=240$ matching the relatively low minijet gluon $dN_g/dy$
in Ref.~\cite{v2} (crosses), and for $dN/dy=645$ matching the roughly $\sim 2.5\times$ higher 
quark density ``string melting'' initial conditions in AMPT (open circles). For a comparison of
rapidity densities, see Fig.~\ref{Fig:comp_dNdy}. As in Ref.~\cite{v2}, MPC was used with massless partons,
constant formation proper time
$\tau_0 = 0.1$~fm, and locally thermal initial momentum distributions with temperature $T_0 = 0.7$~GeV.
}
\label{Fig:comp_v2}
\end{figure}

One might expect that AMPT generates sufficient $v_2$ simply due to its enhanced quark density coming from
string melting. However, as shown in Fig.~\ref{Fig:comp_v2},
the enhanced opacity is still insufficient. The differential elliptic flow%
\footnote{
We define $v_2$ as the average $\cos 2\phi$ in the respective $p_T$ bin,
relative to the theoretical impact parameter direction that is always $\phi = 0$ (the $x$ axis) in the simulation.
}
 of quarks from AMPT (pluses)
plateaus at about $6$\%. In contrast, the covariant transport solution obtained
using MPC with the low minijet $dN/dy=240$ of Ref.~\cite{v2}
gives only $v_2 \approx 1.5$\% (open circles). With opacities increased to match $dN/dy = 645$ from string melting,
elliptic flow only reaches about $3.5$\% at high $p_T$ (crosses).
%

Note that the MPC results in Fig.~\ref{Fig:comp_v2} are for massless quanta, formed at Bjorken proper time $\tau_0 = 0.1$~fm.
The average formation time of quarks in the AMPT calculation is about twice higher,
$\langle \tau \rangle \approx 0.22$~fm.
If quarks are formed at $\tau_0 = 0.22$~fm in MPC, then the initial quark density $n\sim n_\eta/\tau_0A_\perp$ is lower, 
and thus the final elliptic flow is also somewhat smaller. The effect is much weaker than linear in density, e.g.,
a six-fold increase formation time to $\tau_0 = 0.6$~fm results in a $\approx 30$\%  drop in $v_2$ (not shown).

With the simple $1\ {\rm parton} \to 1\ {\rm pion}$ hadronization employed in Ref.~\cite{v2},
the $6$\% elliptic flow from AMPT would, of course, yield only about $6$\% charged hadron $v_2$.
On the other hand, AMPT hadronizes via quark coalescence\cite{Greco:2003xt,Fries:2003vb} instead, which can greatly
amplify elliptic
flow\cite{coalv2}. The coalescence algorithm in AMPT would deserve further scrutiny, 
especially regarding the flow enhancement, but here we focus only on the partonic transport stage.
As seen above, even at the parton level, there is a puzzling factor of two or so discrepancy remaining
between the elliptic flow from AMPT and covariant MPC results at AMPT opacities.

\subsection{Reproducing AMPT events, and covariance}
\label{Sc:repro}

Before isolating additional ingredients in AMPT that help increase elliptic flow,
we test whether AMPT results are reproduced if we feed the partonic initial conditions of AMPT (position, momentum, and formation
time for each quark),
event-by-event, into MPC. Note that, for this cross-check, MPC has to be run without
parton subdivision. So, in effect, this is a comparison of
the noncovariant dynamics in AMPT to the noncovariant $\ell = 1$ dynamics in MPC
using identical initial conditions.

\begin{figure}[h]
\leavevmode
\hspace*{-1mm}
\includegraphics[height=65mm]{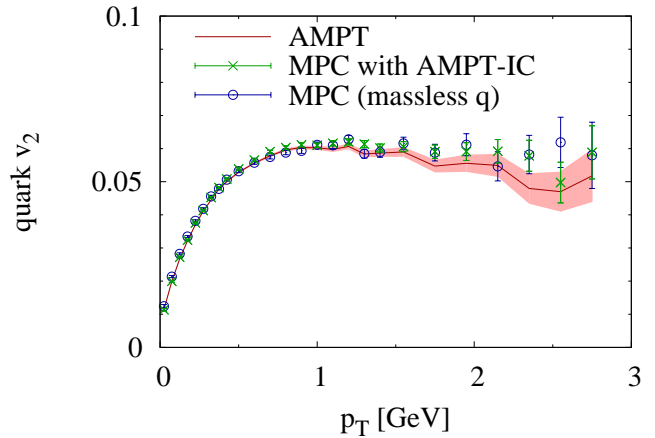}
\caption{Differential quark elliptic flow $v_2$ vs $p_T$ at midrapidity ($|y| < 1$) in Au+Au at RHIC with $b=8$~fm
from AMPT, compared to results from MPC using the same partonic $d\sigma/dt$ 
{\em and also same event-by-event partonic initial conditions} 
(positions, momenta, formation times) as in AMPT, with $\sigma_{qq} = 3$~mb. 
For this comparison, MPC had to be run without parton subdivision ($\ell =1$), i.e.,
these results are {\em not} solutions of the covariant BTE. 
Still, $v_2(p_T)$ from AMPT (solid curve with shaded error band)
is reproduced completely, within statistical errors, by the MPC simulations. 
This is so both for MPC runs with massive quarks (crosses), with
$m_u=9.9$~MeV, $m_d=5.4$~MeV, $m_s=199$~MeV set as in AMPT,
and when massless quarks are used (open circles).
}
\label{Fig:v2_amptIC}
\end{figure}

Figure~\ref{Fig:v2_amptIC} shows the differential elliptic flow of quarks as a function of $p_T$.
The $v_2(p_T)$ result from AMPT (solid line, shaded error bands) is, of course, identical to the one in Fig.~\ref{Fig:comp_v2}.
Within statistical errors, it is reproduced perfectly by the MPC simulations. In fact,
two different MPC simulations are shown: one with massless quarks (open circles),
and the other with quark masses set as in AMPT (crosses). The difference
between the two cases is negligible in $v_2$, 
which is not too surprising because the $u$ and $d$ quarks are practically massless
in AMPT, and the strange quark mass is also quite small.
%

\begin{figure}
\leavevmode
\includegraphics[height=65mm]{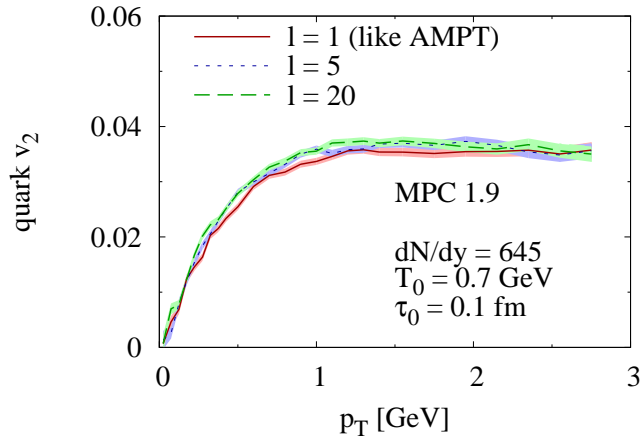}
\caption{Differential elliptic flow $v_2$ vs $p_T$ at midrapidity ($|y| < 1$) in Au+Au at RHIC with $b=8$~fm
from MPC, for different parton subdivision factors $\ell$. Longitudinally boost-invariant initial conditions
with $dN/dy = 645$, formation time $\tau_0 = 0.1$~fm, and locally thermal initial momenta with temperature 
$T_0 = 0.7$~GeV were used. 
The naive cascade $v_2(p_T)$ result with no subdivision 
($\ell = 1$, solid curve) is compared to results with $\ell = 5$ (dotted), and 20 (dashed). 
Shaded bands indicate $\pm 1\sigma$
statistical errors.  }
\label{Fig:v2_vs_l}
\end{figure}

While it is reassuring that the parton dynamics of AMPT matches that of MPC in naive cascade mode (no subdivision),
it raises the question how different such noncovariant solutions are from actual solutions of the covariant BTE (\ref{BTE}).
Unfortunately, it is not possible to answer this question directly because there is no unique way to implement
parton subdivision (i.e., divide particles into multiple test particles) on an event-by-event basis.
Nevertheless, one can quantify noncovariant artifacts on elliptic flow for ''AMPT-like'' initial conditions
while also using the same microscopic cross sections as in AMPT.
Figure~\ref{Fig:v2_vs_l} shows $v_2$ as a function of transverse momentum in Au+Au at RHIC with $b=8$~fm obtained from MPC,
for massless quanta, with string-melting opacities $dN_q/dy = 645$, formation time $\tau_0 = 0.1$~fm, 
initial parton temperature $T_0=0.7$~GeV.
Results for different subdivision factors $\ell = 1$ (solid line), 5 (dotted), and 20 (dashed) are compared,
with statistical errors shown as shaded bands.
The elliptic flow curves practically agree for $\ell = 5$ and $20$, indicating that convergence to the correct
solution of the BTE is reached at $\ell \sim 10$. However, the naive cascade result ($\ell = 1$)
is systematically below the covariant one by $5-10$\%.
We therefore conclude that noncovariant artifacts from the partonic stage of AMPT 
likely give less than $10$\% relative error in $p_T$-differential elliptic flow, 
for string melting initial conditions in Au+Au at RHIC.
%

Thus, the lack of parton subdivision in AMPT has a small effect on $v_2$ calculations at RHIC energies.
This appears to be in stark contrast with the large nonlocal artifacts found in Ref.~\cite{v2}
that only disappeared when high subdivisions $\ell \sim 200$ were used.
The reason why $\ell \sim {\cal O}(100)$ is {\em not} required in AMPT
is that i) AMPT, effectively, incorporates some parton subdivision already,
and ii) the parton system in AMPT is more dilute.
The most opaque calculations in Ref.~\cite{v2} used isotropic $\sigma_{gg} = 20$~mb.
Relative to that, the 3-mb elastic cross section in AMPT corresponds to a subdivision of 
$\ell_{eff} = 20 {\ \rm mb}/3{\ \rm mb} \approx 7$.
In addition, the parton opacity is proportional to the rapidity density and the elastic cross section. 
For the study in Ref.~\cite{v2}, $\sigma_{gg} dN/d\eta = 240 \times 20{\ {\rm mb}} = 4800$~mb;
while for AMPT with string melting, the opacity is $2.5\times$ smaller, 
$\sigma dN/d\eta \approx 645 \times 3{\ {\rm mb}} \approx 1900$~mb.
At such conditions, running ZPC with subdivision $\ell \sim 10$ 
would indeed be sufficient to make the partonic stage of AMPT practically covariant.

\subsection{Influence of initial momentum distribution}
\label{Sc:mom}

The results in Sec.~\ref{Sc:repro} establish
that the parton dynamics in AMPT is fairly close to solutions of the covariant BTE (for RHIC collisions at least),
and yet, as seen in Sec.~\ref{Sc:prof}, the initial parton $dN/dy$ is insufficient
to generate as high $v_2$ as AMPT does with its $2\to 2$ partonic cross section.
The origin of the substantial elliptic flow must then lie in other aspects of the AMPT initial conditions.
In fact, the key missing ingredient turns out to be the initial momentum distribution.

\begin{figure}[h]
\leavevmode
\includegraphics[height=65mm]{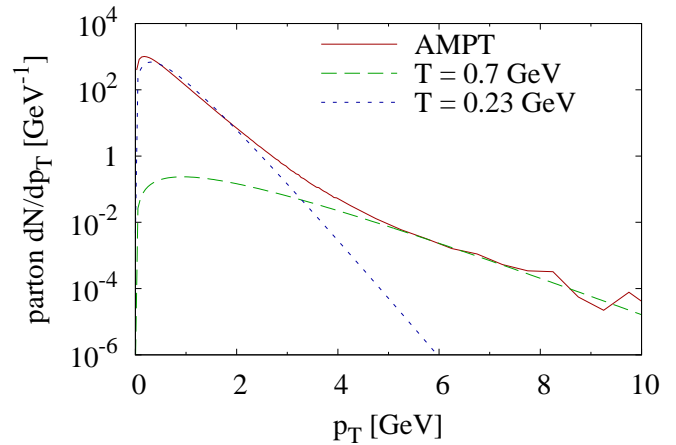}
\caption{Initial quark momentum distribution $dN/dy dp_T$ 
at midrapidity ($|y|<1$) in Au+Au at RHIC with $b=8$~fm from AMPT (solid line).
The high-$p_T$ tail of the distribution is well characterized by an effective temperature $T_{eff}=0.7$~GeV (dashed),
the same value as the initial minijet temperature used in Ref.~\cite{v2}.
However, the bulk of the AMPT initial quark distribution corresponds to much lower $T_{eff} \approx 0.23$~GeV (dotted).
Both thermal curves are shown for massless partons.
}
\label{Fig:pt_dist}
\end{figure}

Figure~\ref{Fig:pt_dist} shows the initial transverse momentum distribution $dN/dp_T dy$ of quarks at midrapidity 
in Au+Au at RHIC at $b=8$~fm from AMPT (solid line). The high $p_T \gton 4$~GeV tail of the distributions
is close to a local thermal distribution with effective temperature $T_0 = 0.7$~GeV (dashed),
which is the same as the minijet temperature used to initialize gluons in Ref.~\cite{v2}.
This temperature, however, is not representative for the vast majority of quarks in AMPT. For example,
for the longitudinally boost-invariant thermal distribution (\ref{thermal_boostinv}), the average transverse
momentum of particles is $\langle p_T\rangle = 3\pi T_0/4$, about $1.65$~GeV at $T_0 = 0.7$~GeV,
whereas quark in AMPT have $\langle p_T\rangle \approx 0.54$~GeV at formation.
Thus, the bulk of the system in AMPT has a three times lower effective temperature 
$T_0 \approx 0.23$~GeV (dotted)%
\footnote{This value is in qualitative agreement with $T\sim 0.2-0.3$~MeV extracted
from the center of the AMPT collision zone by an earlier analysis in Ref.~\cite{Lin:2014tya}.
}
This profoundly affects the transport opacity in the system, and thus elliptic flow as well.
%

Elliptic flow, of course, depends on the elastic cross section.
But, as shown in Ref.~\cite{v2}, what really matters is the transport
opacity $\chi \sim \int dz \sigma_{tr} n$. Here, the transport cross section
\be
\sigma_{tr} \equiv \int d\sigma \sin^2 \theta_{cm} 
= \int dt \frac{d\sigma}{dt} \frac{4t}{s}\left(1 - \frac{t}{s}\right)
\ee
weights large-angle scatterings preferentially.
For the screened, perturbative cross section (\ref{dsigmadt}) in AMPT,
\be
\frac{\sigma_{tr}}{\sigma_{tot}} = 4 z (1+z) \left[(2z + 1) \ln \left(1 + \frac{1}{z}\right) -2 \right] 
\label{sigmatr}
\ee
is a monotonically increasing function of $z \equiv \mu^2/s \approx \mu^2 / 18T^2$,
where we substituted $\langle s\rangle = 18T^2$ for a thermal system of massless partons.
Thus, the smaller the Debye mass to temperature ratio, the less efficient scatterings are
for generation of elliptic flow.
At the minijet temperature $T_0 = 0.7$~GeV, $\mu / T_0 \approx 0.64$, so $\sigma_{tr} \approx 0.18 \sigma_{tot}$. 
In contrast, at the effective temperature $T_0 \approx 0.23$~GeV of AMPT quarks, $\mu / T_0 \approx 1.95$, so
$\sigma_{tr} \approx 0.5\sigma_0$. This more than two-fold increase in the transport opacity due to the rather 
low initial quark temperature (and hence more isotropic scatterings) 
is the missing link needed to explain the high $v_2$ results from AMPT.

\begin{figure}[h]
\leavevmode
\includegraphics[height=65mm]{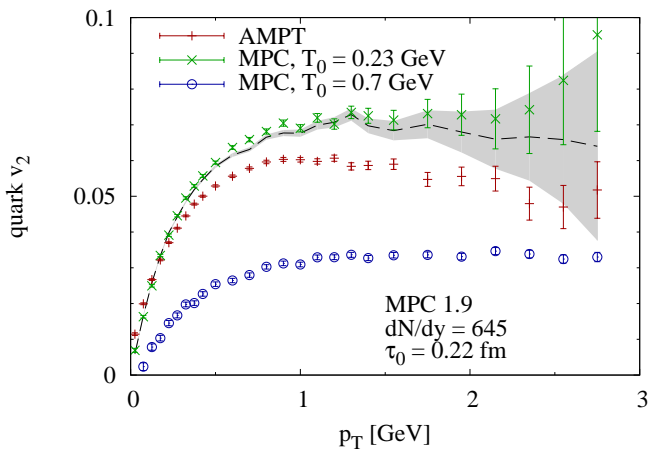}
\caption{Differential quark elliptic flow $v_2$ vs $p_T$ at midrapidity ($|y| < 1$) in Au+Au at RHIC with $b=8$~fm
from AMPT (pluses), compared to covariant results from MPC using the same partonic 
$d\sigma/dt$ as in AMPT with $\sigma_{qq} = 3$~mb. MPC calculations from longitudinally boost-invariant initial
conditions are shown for massless partons with locally thermal initial momenta at
temperature $T_0 = 0.7$~GeV (open circles), as for minijet gluons in Ref.~\cite{v2}, 
and three times lower $T_0 = 0.23$~GeV (crosses) to match the initial quark $\langle p_T \rangle$ in AMPT.  
The rapidity density $dN_q/dy=645$ and formation time $\tau_0 = 0.22$~fm in MPC were set
to the peak $dN_q/dy$ and average formation time for ``string melting'' initial conditions in AMPT. 
Covariant MPC results with $T_0 = 0.23$~GeV are also shown for formation times generated randomly according to the 
formation time distribution $dN_q/d\tau$ from AMPT (dashed black line with shaded error bands), instead
of using the average value $\tau_0 = \langle \tau \rangle = 0.22$~fm.
}
\label{Fig:v2_vs_Teff}
\end{figure}

To demonstrate the final resolution to the AMPT opacity puzzle, we
present in Fig.~\ref{Fig:v2_vs_Teff} covariant MPC calculations for the differential elliptic flow $v_2(p_T$) 
of massless quarks in Au+Au at RHIC,
using string-melting initial quark density $dN_q/dy = 645$, AMPT's average formation time $\tau_0 = 0.22$~fm,
and two different initial temperatures $T_0 = 0.23$ and $0.7$~GeV.
With quarks initialized at the $0.7$~GeV minijet temperature (open circles), elliptic only reaches up to about $3$\%
(almost the same as the $v_2$ for $\tau_0 = 0.1$~fm shown in Sec.~\ref{Sc:prof} (Fig.~\ref{Fig:comp_v2}, crosses)).
On the other hand, with $T_0 = 0.23$~GeV from string melting, elliptic flow reaches up to $7$\% (crosses),
so it even exceeds slightly the flow from AMPT (pluses).
Figure~\ref{Fig:v2_vs_Teff} also shows the most AMPT-like covariant calculation in this work 
(dashed line, shaded error bands),
with not only the high rapidity density and low effective temperature of string melting initial conditions 
in AMPT
but also quark formation times generated according to the formation time $dN/d\tau$ distribution in AMPT.
The latter results are practically the same as those obtained with the average formation time $\tau_0 = 0.22$~fm.

\section{Discussion and conclusions}

In this work we explain how the widely used AMPT transport model\cite{AMPT} generates
sufficient elliptic flow in A+A reactions with only 3-mb elastic cross sections in its early
partonic stage.
The key ingredients needed are pinpointed through detailed comparisons to covariant Boltzmann transport
solutions obtained with Molnar's
transport code MPC\cite{Molnar:2000jh,MPCcode}, in which the simplified conditions studied in Ref.~\cite{v2}
are gradually revised to more and more resemble those in AMPT.
We focus on the partonic stage of AMPT, and 
only study Au+Au collisions at $\sqrt{s_{NN}} = 200$~GeV with impact parameter $b=8$~fm
($\approx 30$\% centrality), 
but the findings are expected to hold quite generally.

Two features are primarily responsible for the high AMPT $v_2$ in Au+Au
at RHIC (cf. Secs.~\ref{Sc:prof} and \ref{Sc:mom}):
i) a high initial parton rapidity density $dN/dy \approx 650$ from the ``string-melting'' 
scenario\cite{AMPTsm}, 
about $2.5\times$ higher than the
minijet $dN/dy \approx 240$ studied in Ref.~\cite{v2};
and ii) an initial effective temperature $T_0 \approx 0.23$~GeV that is rather "cold" in the perturbative sense,
about $3\times$ lower than the 0.7~GeV minijet temperature used in Ref.~\cite{v2}. 
These findings are in qualitative agreement with Ref.~\cite{Lin:2014tya}, which showed
that quarks in AMPT are in fact oversaturated (their fugacity is above one).
The higher Debye mass over temperature ratio $\mu/T \approx 2$ leads to much more isotropic scatterings,
and hence more efficient generation of elliptic flow. 
This is reminiscent of the picture in Ref.~\cite{Liao:2006ry},
where the onset of chromomagnetic monopoles provides a big increase in interactions and 
much more efficient transport at fairly
low temperatures $T \sim 100-200$~MeV, in the vicinity
of the QCD transition temperature. Albeit AMPT starts pretty much in the cold phase already,
instead of having a high-temperature stage, subsequent cooling, and then sudden onset of microscopic activity.

We also address the concern that the parton dynamics in AMPT is not covariant
because AMPT does not employ parton subdivision\cite{Zhang:1998tj,Molnar:2000jh}.
With the help of the covariant MPC code, we estimate that noncovariant artifacts 
distort differential elliptic flow
$v_2(p_T)$ for quarks in Au+Au at RHIC from AMPT by less than $10$\%, i.e., such errors are modest
(Sec.~\ref{Sc:repro}).
In addition, we demonstrate that MPC in naive cascade mode (which does not use subdivision) reproduces the quark
$v_2(p_T)$ from AMPT perfectly if one runs it with partonic initial conditions taken, event-by-event, from AMPT.
This gives one confidence that AMPT results are relatively close to covariant transport solutions.

Even with its high parton rapidity density and near-isotropic scatterings,
the quark $v_2(p_T)$ from AMPT only reaches about $6$\%,
which is far below the measured hadron $v_2$.
AMPT makes up for the difference by hadronizing via quark coalescence\cite{Greco:2003xt,Fries:2003vb}.
For quarks that are comoving, it has been shown\cite{coalv2} that coalescence
indeed amplifies $p_T$-differential elliptic flow $2-3\times$ compared to that of quarks.
(The fusion process of coalescence should also help with getting enough $\langle p_T \rangle$ for hadrons from
the ``cold'' AMPT quark plasma.)
On the other hand, it is not clear how close coalescence dynamics in AMPT
is to the comoving limit. Therefore, it would be very worthwhile to investigate the 
coalescence algorithm in AMPT in detail in the future. 

{\it Acknowledgments.}
Helpful discussions with Z.W.~Lin and M.~Gyulassy are acknowledged. 
This research was supported in part by the US Department of Energy, Office of Science, under
Award No. DE-SC0016524; and through computational resources provided by Information Technology at Purdue, 
West Lafayette, Indiana.


\end{document}